\documentclass[aps, prb, showpacs,preprint]{revtex4-1}

\usepackage{color}

\pdfoutput=1
\usepackage{graphicx}

\pdfoutput=1
\begin{document}
\title{Jahn-Teller distortion induced magnetic phase transition in cubic BaFeO$_{3}$}

\author{Gul Rahman$^{1}$}\email{gulrahman@qau.edu.pk}
\affiliation{$^1$Department of Physics,
Quaid-i-Azam University, Islamabad 45320, Pakistan}

\author{Juliana M. Morbec$^{2,3}$}
\affiliation{$^2$Department of Chemistry, University of California, Davis, California 95616, USA}
\affiliation{$^3$Instituto de Ci\^encias Exatas, Universidade Federal de Alfenas, 37130-000, Alfenas, MG, Brazil}

\author{Rub\'en Ferrad\'as$^{4,5}$}
\affiliation{$^4$ Departamento de F\'{\i}sica, Universidad de Oviedo, 33007 Oviedo Spain}
\affiliation{$^5$ Nanomaterials and Nanotechnology Research Center (CINN), Spain}

\author{V\'{\i}ctor M. Garc\'{\i}a-Su\'arez$^{4,5,6}$}
\affiliation{$^4$ Departamento de F\'{\i}sica, Universidad de Oviedo, 33007 Oviedo Spain}
\affiliation{$^5$ Nanomaterials and Nanotechnology Research Center (CINN), Spain}
\affiliation{$^6$ Department of Physics, Lancaster University, Lancaster LA1 4YW, United
Kingdom}

\begin{abstract}
Using density functional theory (DFT) with local density approximation (LDA) and generalized gradient approximation (GGA) correlation functionals, the electronic and magnetic structures of cubic BaFeO$_{3}$ in the ferromagnetic (FM) and
antiferromagnetic (AFM) states are studied. Our LDA/GGA and LDA$+U$/GGA$+U$ results show that
cubic BFO has a FM ground state, in agreement with recent experimental works. Two types of
Jahn-Teller (JT) distortions, denoted as JT1 and JT2, are considered. We find FM to
ferrimagnetic (FIM) and FM to AFM magnetic phase transitionn in the JT1 and JT2 type of
distortions, respectively. Larger strains are required for the FM-AFM transition as compared to
the FM-FIM. DFT$+U$ calculations also show that the magnetic moments dramatically decrease
at large strains due to strong overlapping between the Fe and O atoms. The origins of these
transitions is discussed in terms of a competition between double exchange and
superexchange interactions. Oxygen and Fe displacements are therefore responsible for the
magnetic phase transitions and the reduction of the magnetic moments. 
\end{abstract}


\maketitle

\section{Introduction}
\label{sec:intro}
Iron-based oxides have very interesting properties due to the different oxidation states of Fe,
which give rise to different crystal structures and stoichiometries. Well known examples are
FeO and Fe$_{3}{\rm{O}}_{4}$ where Fe can exist in Fe$^{2+}$ and Fe$^{2+/3+}$ states,
respectively. All these materials are antiferromagnetic (AFM) or ferrimagnetic (FIM)
insulators. Fe-containing ferroelectric oxides are currently attracting tremendous attention, not only due to their magnetic properties but also due to their applications. Particularly, those oxides that exhibit magnetic and ferroelectric characteristics simultaneously, known as
multiferroics, can have practical device applications such as spin transistor memories, whose magnetic properties can be tuned by electric fields through the lattice strain effect.\cite{ref1} A lot of work has been done on perovskite  materials to understand the origins of multiferroic behavior.~\cite{ref2} Perovskite  materials containing Fe, e.g., BiFeO$_{3}$\cite{bifeo3} and SrFeO$_{3}$ (SFO)~\cite{ref3,ref4}, have very complicated magnetic structures that promote such type of behaviors.\cite{HMnO3,GFO}
In cubic SFO, which is AFM and metallic down to low temperatures\cite{ref3,ref4}, Fe is surrounded by six O atoms and it is in the Fe$^{4+}$ state. SFO is of particular interest due to its simple crystal structure and helical magnetic structure. Recently, the helical spin order in Fe perovskites has been investigated using a double exchange (DE) model including oxygen $2p$ orbitals.\cite{Most} In particular, it was found that G-type helical magnetic structure could be changed to A-type by reducing the superexchange (SE) interactions.~\cite{Most}

Another example of perovskites is BaFeO$_{3}$ (BFO), which is one of the few oxides where iron has an oxidation state of +4.\cite{Lucchini}
In bulk, BFO normally assumes a hexagonal crystal structure, although various
polymorphs have been observed with oxygen deficiency.\cite{Lucchini,Hook,Hombo,Iga,WW}
Bulk hexagonal BFO exhibits an interesting AFM to FM transition at 160 K.\cite{Mori}
Recent experiments observed however ferromagnetism in cubic BFO, where the metastable cubic structure was shown to be stable down to $8$K,\cite{expt} and a field-induced switching from AFM to FM. BFO has also spin spiral magnetic order, but it is different from that of SFO. As a consequence, SFO needs larger external fields than BFO for the AFM to FM transition.
Callender {\it et al.}\cite{Calender} have grown epitaxially cubic BFO on STO and found week ferromagnetism with a transition temperature of 235 K.
Similarly, Matsui {\it et al.}~\cite{Matsui} have also reported ferromagnetic (FM) in pseudocubic BFO on STO.
Very recently, Li {\it et al.}\cite{LiPRB2012} used DFT$+U$ to investigate the helical structure of cubic FM BFO and SFO and found AFM to FM transitions under external pressure. The structural FM stability of cubic BFO has also been studied experimentally and it was found that cubic BFO remains FM up to 40 GPa.\cite{Naoaki} The magnetic structure of BFO can also be tuned by external small magnetic field (0.3 T).\cite{Naoaki} Such experimental and theoretical results indicate that magnetic phase transitions (AFM to FM) can easily be affected by external perturbations. Motivated by this recent experimental work on cubic BFO, here we demonstrate that this BFO structure shows FM to AFM/FIM phase transitions due to Jahn-Teller(JT) type distortions, which can be used to engineer BFO for applications.

\section{Computational Method}
We performed calculations in the framework of density functional theory (DFT) ~\cite{DFT} using linear combinations of atomic orbitals (LCAO) as implemented in the SIESTA code ~\cite{siesta}. We used a double-$\zeta$ polarized (DZP) basis set for all atoms. We employed both the local density approximation (LDA) as parametrized by Ceperly and Adler, \cite{lda} and  the generalized gradient approximation (GGA) as parametrized by Wu and Cohen \cite{WC} for the exchange-correlation functional.
We used standard norm-conserving pseudopotentials ~\cite{ps} in their fully nonlocal form .~\cite{pss} We carefully checked the convergence of the cutoff energy for the real-space grid, since the lattice parameters and magnetic moments in this system were rather sensitive to this parameter. This energy cutoff defines the energy of the most energetic plane wave that can be represented on such grid, i.e., the larger the cutoff the smaller the separation between points in the grid ($E\sim G^2\sim 1/d^2$, where $\vec G$ is a reciprocal vector and $d$ is the separation between points). We found that 400 Ry was enough to converge the lattice parameters and magnetic moments. Similarly, we performed the
Brillouin zone integrations using different Monkhorst-Pack grids. We found that the magnetic moments and lattice parameters were converged with a $24\times 24\times 24$ mesh. With such optimized computational parameters, we obtained $3.88$ and $3.94$\,\AA\, for the the LDA and GGA lattice constants $a$, respectively, which are comparable to the experimental lattice constant, $3.99$ \AA\cite{verma} or $3.97$ \AA. \cite{Naoaki} Using such optimized lattice constants, we obtained magnetic moments per Fe atom of 3.40 $\mu_{\rm{B}}$ and 3.58 $\mu_{\rm{B}}$ for LDA and GGA, respectively, which are also comparable to the experimentally determined magnetic moment $\sim 3.50$ $\mu_{\rm{B}}$.~\cite{Naoaki}

We performed the LDA$+U$ and GGA$+U$ calculations with the LDA and GGA calculated lattice constants. In order to determine the influence of different $U$'s and cutoff radii and
see how the results compare to previous experiments and calculations, we studied first
the typical case of bulk oxide FeO, \cite{Parmigiani99} where DFT is known to give
qualitatively different behavior (metallic instead of insulating character)\cite{Cococcioni05}.
We performed calculations with $U=4$ eV and $U=4.5$ eV, which is the range of values most
used in the literature for iron \cite{SolovyevDederichs94,Cococcioni05}, and used projectors
with different radii. The parameters that best fitted the experiments and previous calculations
for FeO were, $U=4.0$ eV and $r_{\rm c} = 5.5$ Bohr for GGA$+U$, and $U=4.5$ eV and $r_{\rm c} = 5.0$ Bohr for LDA$+U$, which gave a gap of  approximately  $2.4$ eV.
This value is in excellent agreement with the previous value of 2..4 eV given in Ref. \onlinecite{Parmigiani99}. The quantitative and qualitative agreement of our LDA/GGA and LDA$+U$/GGA$+U$ results with previous theoretical and experimental works guarantees the
computational approach used in the present study.

\section{Result and Discussion}
\label{sec:result}
\begin{figure}[!h]
\includegraphics[width=0.4\textwidth]{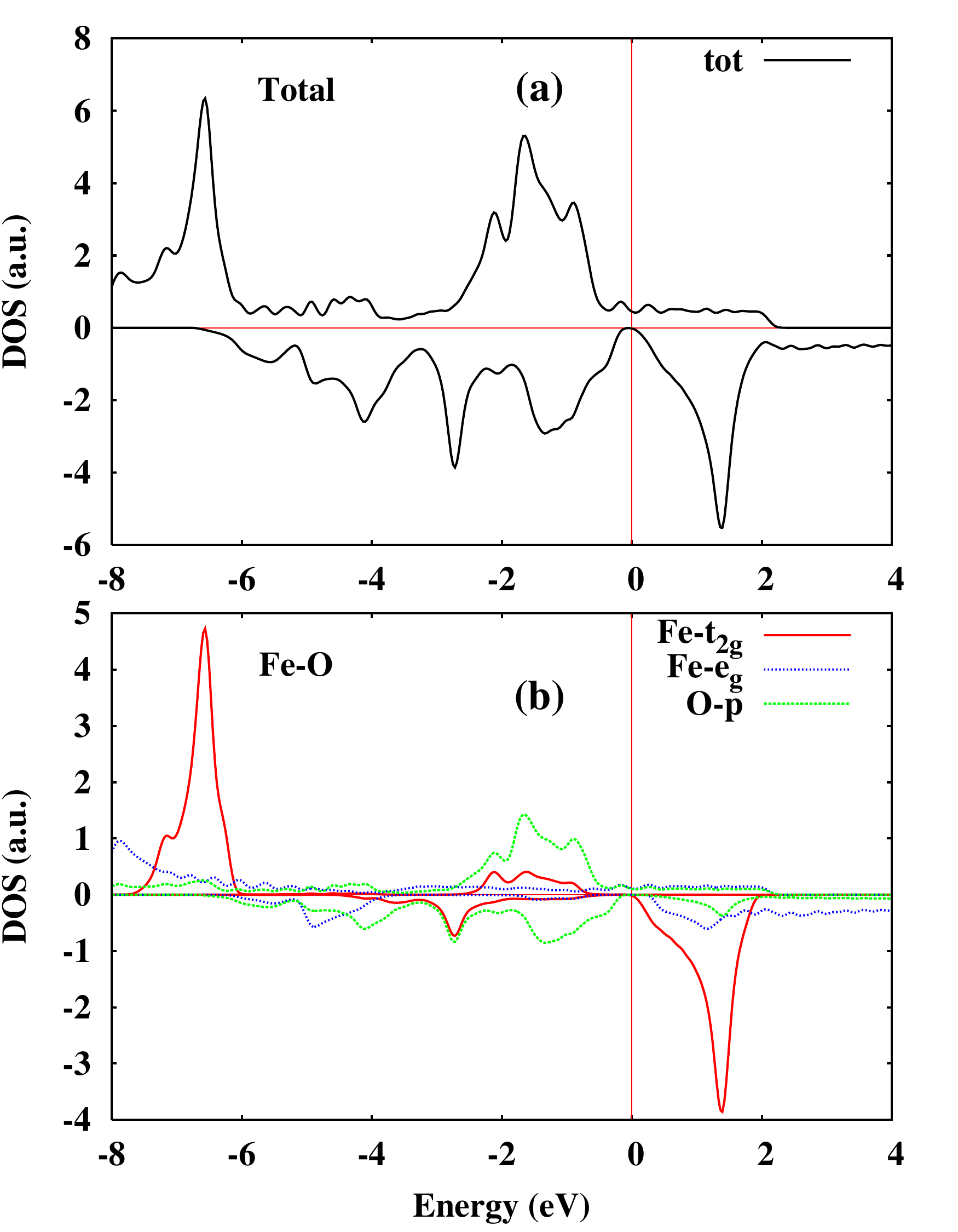}
\caption{(Color online) The calculated total (a) and partial (b) densities of states (PDOS)(in arbitrary units) in the FM states of BFO. Solid (red), dotted (blue), and dashed (green) lines represents the Fe-$t_{2g}$, Fe-$t_{eg}$, and O-$p$ orbitals respectively. The total DOS is represented by solid thick (black) lines. The Fermi energy is set to zero.}
\label{DOSFM}
\end{figure}

We carried out total energy calculations of cubic BFO in nonmagnetic (NM), FM, and AFM
states. We will only discuss GGA$+U$ data, but comparison with LDA$+U$/LDA/GGA will be made
wherever necessary. We found that at ambient pressure BFO has a FM ground state.
Figure~\ref{DOSFM} shows the calculated total and projected density of states (PDOS). The electronic structure of BFO shows typical half metallic behavior, which has important
applications in spin based electronic devices (spintronics).
Strong bonding of Fe $t_{2g}$-O ${p}$ orbitals
can be seen around the Fermi energy both in the majority and minority spin states. The $t_{2g}$-O ${p}$ hybridization is essential for ferromagnetism and can be perturbed either by external pressure or by distorting the local environment around the Fe atoms. Two sharp peaks, contributed by $t_{2g}$ electrons, in the valence spin-up and conduction spin-down states can be seen. This calculated electronic structure is agreement with the recent work of Li {\it et al.}~\cite{LiPRB2012}
It is important to mention that our calculated LDA/GGA electronic structures showed metallic behavior, which is also in agreement with the previous DFT calculations.~\cite{Feng} We therefore believe that including an optimum value of $U$ is essential to describe the true electronic structure of BFO.

\begin{figure}[!h]
\includegraphics[width=0.4\textwidth]{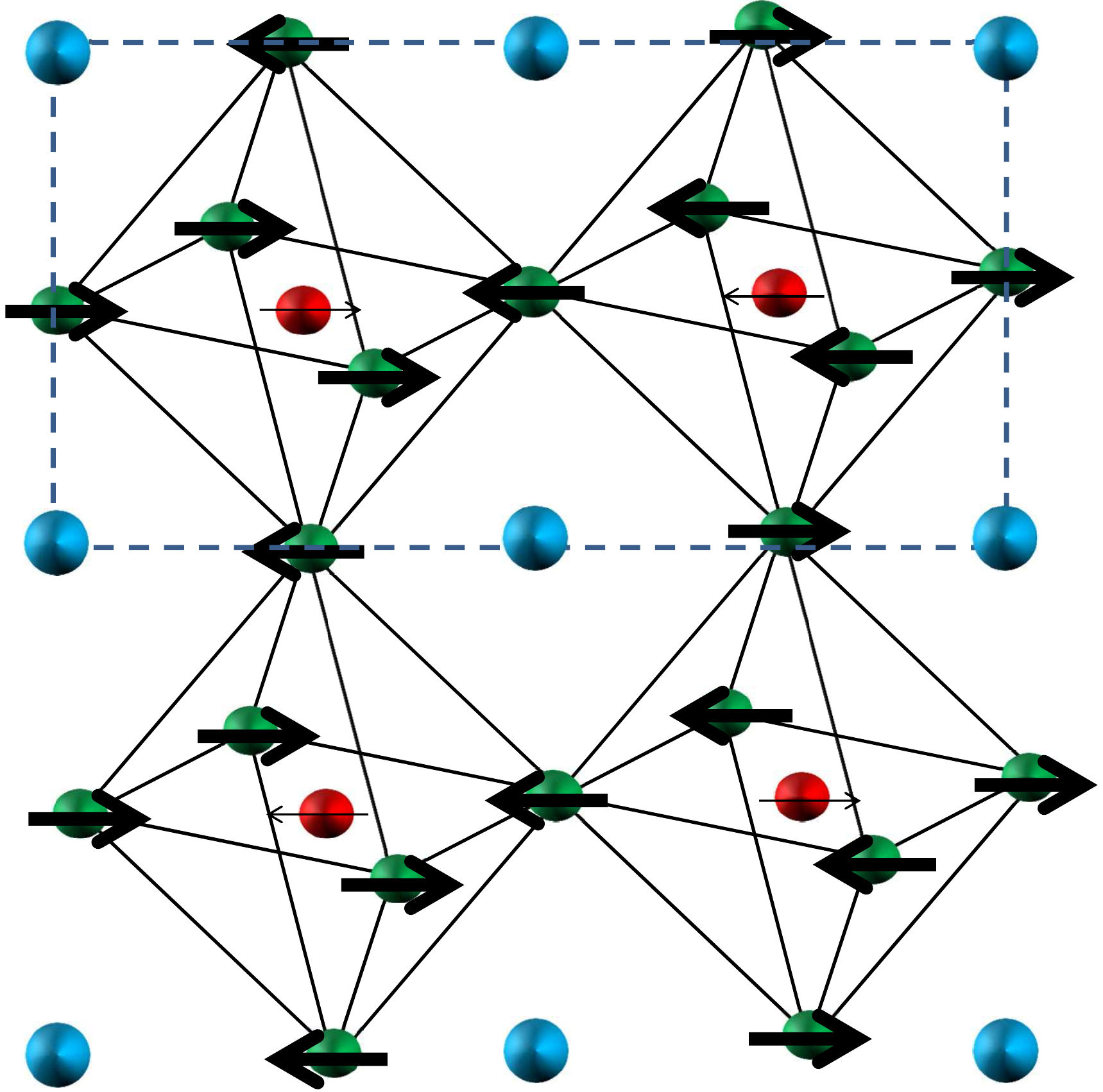}
\caption{(Color online) Schematic representation of JT1 and JT2 distortions. Black thick arrows on the oxygen atoms show the direction of the distortion $\delta$ along $z$ axis.
Red (green) balls show Fe (O) atoms, whereas blue balls show Ba atoms. The dashed lines show that unit cell that is used in the calculations. Thin arrows show the direction of magnetic moments.  Fe$_{\rm I}$ (Fe$_{\rm II}$) has atomic fractional coordinates $0.5,0.5,0.25$ ($0.5,0.5,0.75$ ) }

\label{JT1}
\end{figure}

We mainly considered two types of distortion $\delta$ denoted as JT1 and JT2. In JT1 (see Fig.~\ref{JT1}), the atomic position of the Fe atom is fixed at the center of the cell, whereas the six O atoms surrounding Fe are distorted by $\delta=$ $\{$0, 0.01, ..., 0.10$\}$ \AA\ along the $z$ direction in such a way that the elongation of some of them is compensated by the contraction of the other. We denote these systems as $sys0$, $sys0.01$, ..., $sys0.10$. A double unit cell ($1\times 1\times 2$) was used to take into account AFM coupling between the Fe atoms as well. Notice BFO has BaO and FeO$_{2}$ layers. The two Fe atoms in Fig.~\ref{JT1} are bonded through the O atoms of a BaO layer. We denote these oxygen atoms as O$_{\rm c}$. In the ideal case (no distortion), the in-plane bond angle O-Fe-O is 90, the out of plane bond angle Fe-O$_{\rm c}$-Fe is 180 and the bond lengths is $\sim$ $1.94$ \AA. The distortion of the octahedron around the Fe atoms directly affects the  bond lengths
  and bond angles between Fe and the oxygens.
Since different O atoms are moving in opposite direction, it is expected that consecutive Fe atoms will have different bond lengths and therefore there will be two different kinds of Fe atoms, i.e., Fe$_{\rm I}$ and Fe$_{\rm II}$ (see Fig~\ref{JT1}).
In the second type of distortion JT2 (Fig.~\ref{JT1}), the atomic positions of the O atoms are fixed and the Fe atoms are displaced along the $z$ direction by $\delta$ from its ideal positions, i.e., ($0.5, 0.5, 0.50+\delta$). Notice that in both types of distortions the bond lengths and bond angles are perturbed but the atomic volume is conserved.

\begin{figure}[!h]
\includegraphics[width=0.45\textwidth]{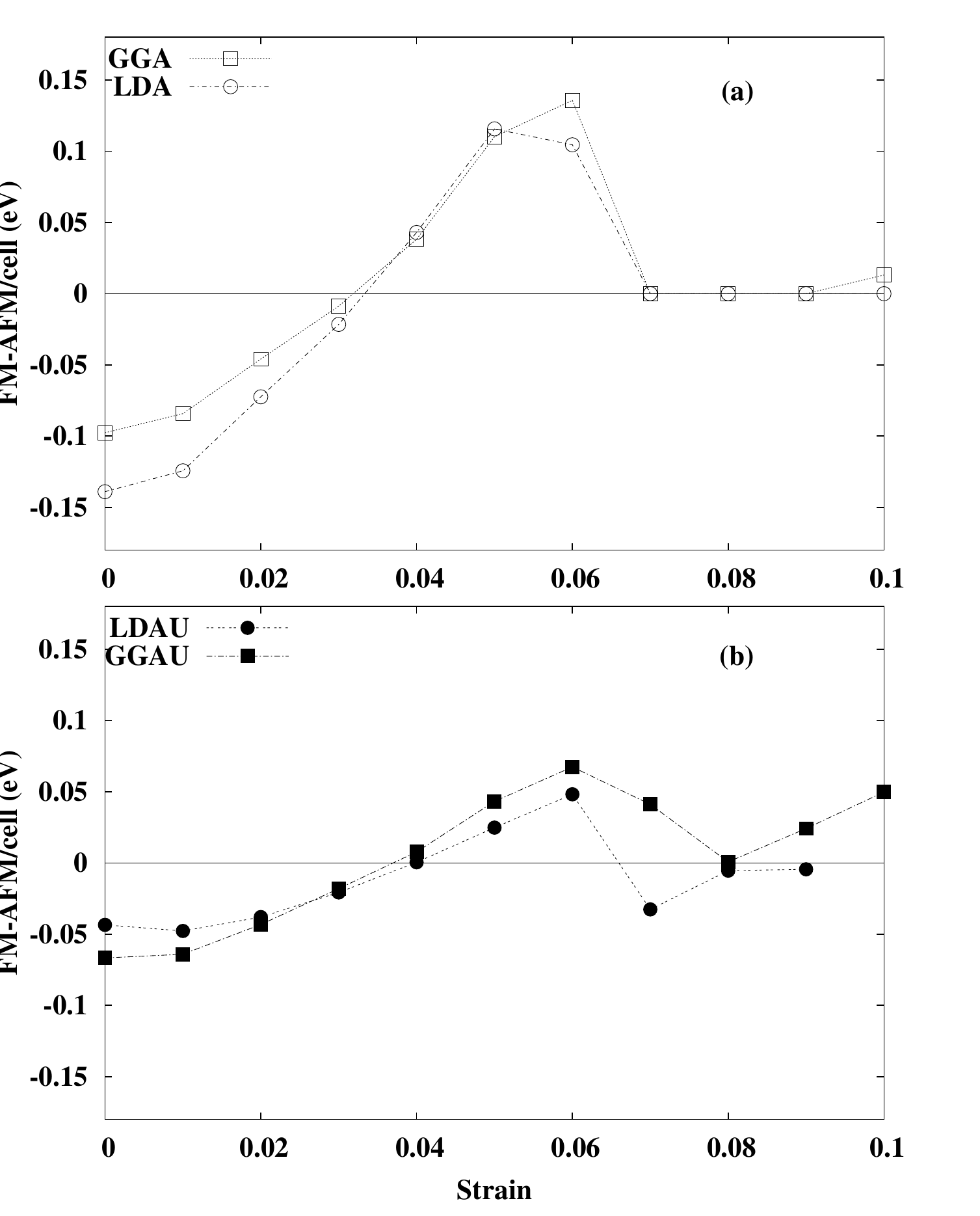}
\includegraphics[width=0.45\textwidth]{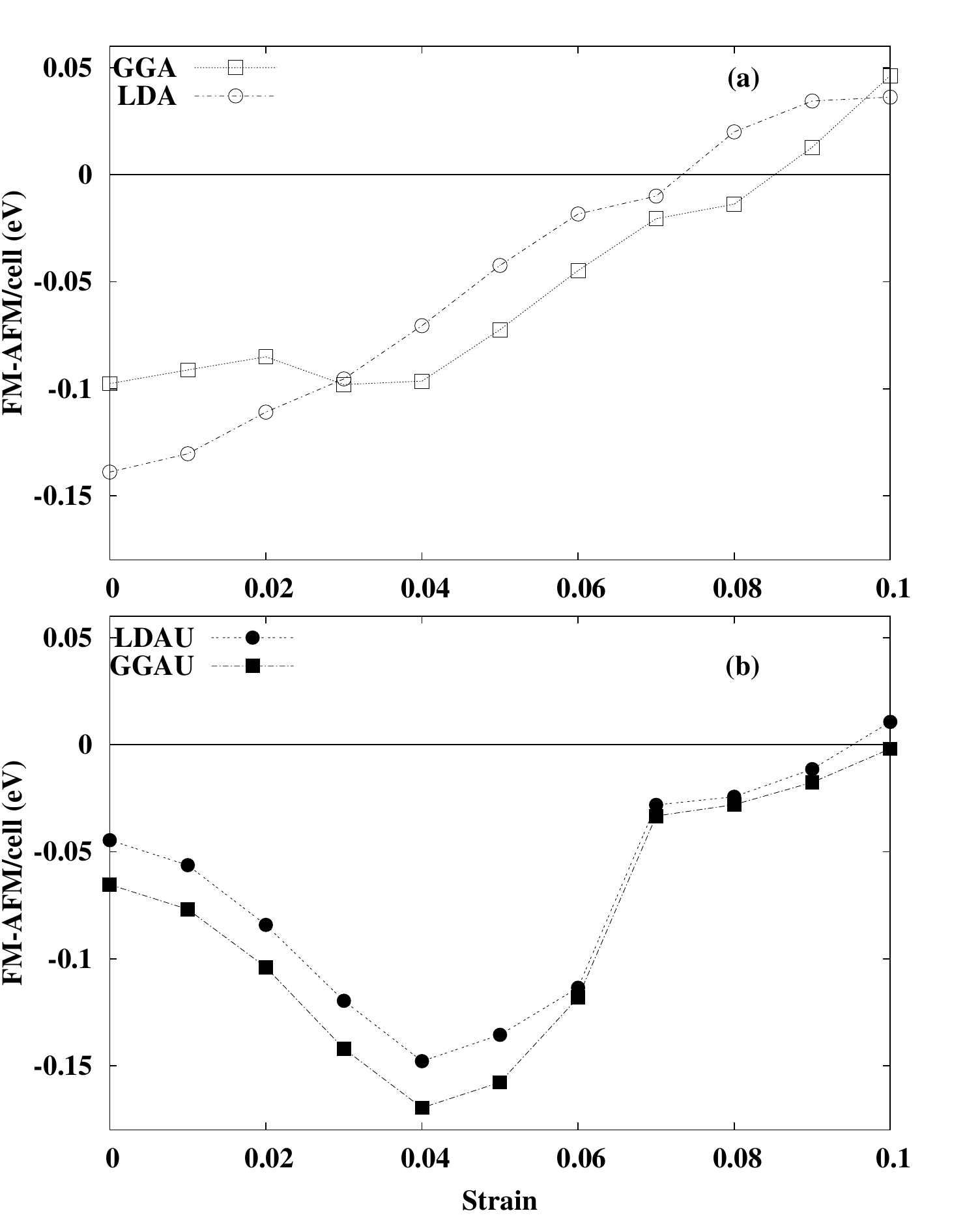}
\caption{(Color online) Total energy difference (FM-AFM)/cell (in eV) vs. strain  for JT1 (left panel) and JT2 (right panel) systems calculated with (a) GGA/LDA and (b) GGA$+U$/LDA$+U$. Empty (filled) squares represent GGA (GGA$+U$) data whereas empty (filled) circles represent LDA (LDA$+U$) results.}
\label{JT1-TE}
\end{figure}

The total energy was calculated for each distortion in the FM and AFM states. The energy difference between the FM and AFM states ($\Delta E=E_{\rm AFM}-E_{\rm FM}$) gives an approximate value of the exchange integral $J$. In the following we will show how $J$ changes with $\delta$ and the exchange correlation functional (without and with $U$).
The LDA/GGA and LDA$+U$/GGA$+U$ calculated $\Delta E$ of JT1 is shown in the left panel of Fig~\ref{JT1-TE}. Negative (positive) $\Delta E$ corresponds to FM (AFM) configurations. In both LDA and GGA the unstrained BFO is in the FM state but as the strain increases, $\Delta E$ decreases and changes sign around $0.03$ eV/\AA$^3$. Such reversal of sign is an indication of  magnetic phase transition, FM $\rightarrow$ AFM. BFO remains in the AFM state in the strained region (0.03 to 0.07) eV/\AA$^3$. Beyond 0.07 eV/\AA$^3$, BFO has however zero $\Delta E$, i.e., FM and AFM are isoenergetic. These calculations demonstrate that  BFO has to cross an energy barrier $\sim 0.1$ eV per cell to transit from one magnetic state (FM) to another magnetic state (AFM). When the calculations were repeated with LDA$+U$/GGA$+U$, a similar magnetic phase transition was obtained. Including $U$, however, decreases the barrier height ($\sim 0.05 $ eV), shifts the phase transition region to higher strains and
  reduces the isoenergetic region.

The second type of strained BFO exhibits a different behavior (see right panel of Fig.~\ref{JT1-TE}). The LDA/GGA calculations show that BFO retains its FM configuration up to 0.07 eV/\AA$^3$ but beyond that point it transits to an AFM state. This shows that large strains would be required for such magnetic phase transition. The phase transition is shifted to even higher strained regions $~ 0.10$ eV/\AA$^3$ when $U$ is included. Interestingly, $\Delta E$ has a local minimum around $0.04$ eV/\AA$^3$, which is exactly the amount of strain where a FM-AFM transition was observed in the JT1 case.
Comparing JT1 and JT2, we see that when some of the oxygen atoms are moved away from the Fe atoms distorting the octahedra the Fe atoms couple antiferromagnetically beyond some strain limit. From this it is possible to infer that O displacements suppress DE and favor SE beyond some critical strain. Changes in the exchange integral $J$ due to strain will therefore have significant effect on DE and SE interactions, as will be discussed below. Moving the Fe atoms away from the oxygen atoms brings however the FM order to lower energy states. These calculations suggest that small atomic displacements are crucial for the magnetic ground state structure.

\begin{figure}[!h]
\includegraphics[width=0.4\textwidth]{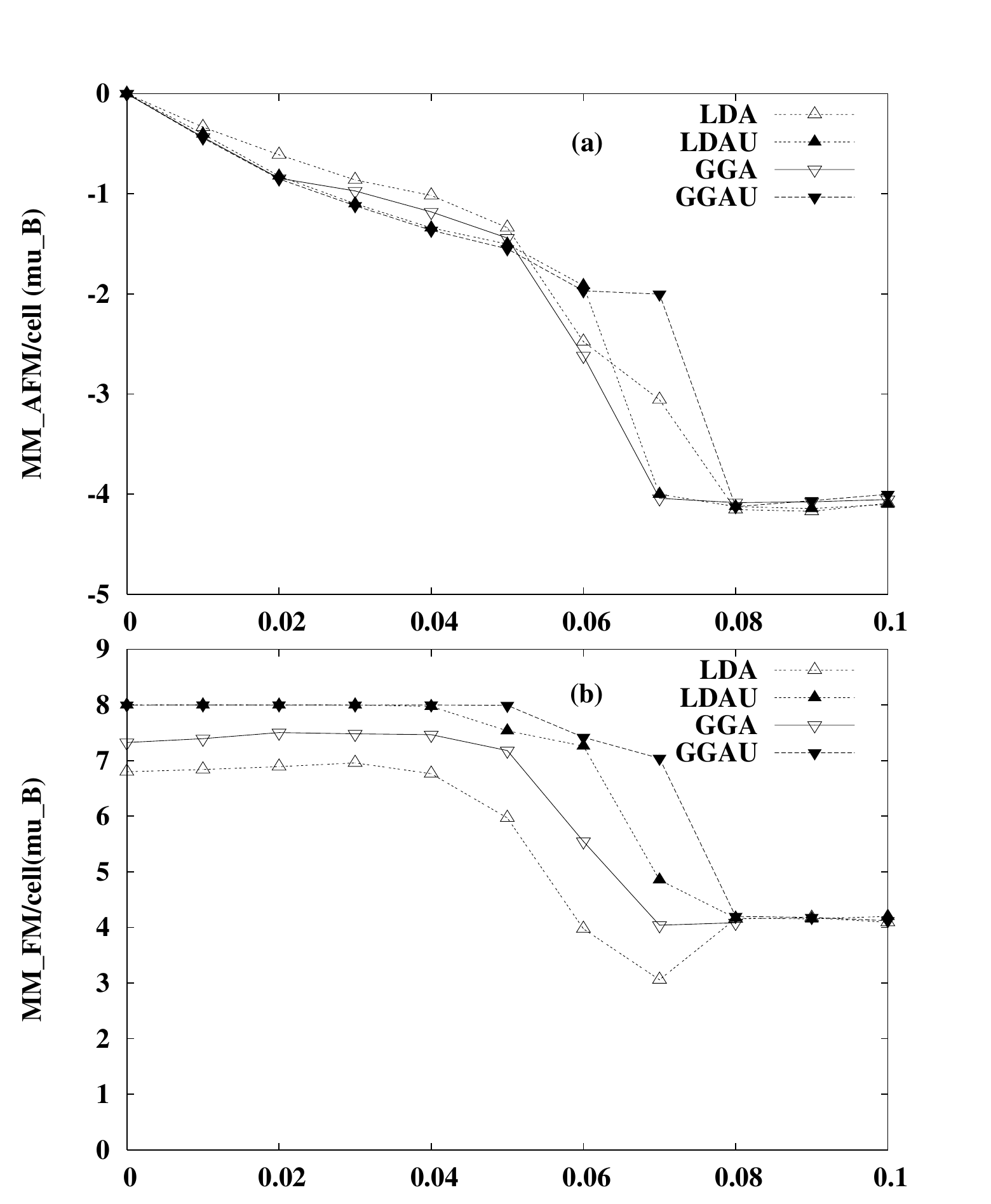}
\includegraphics[width=0.32\textwidth]{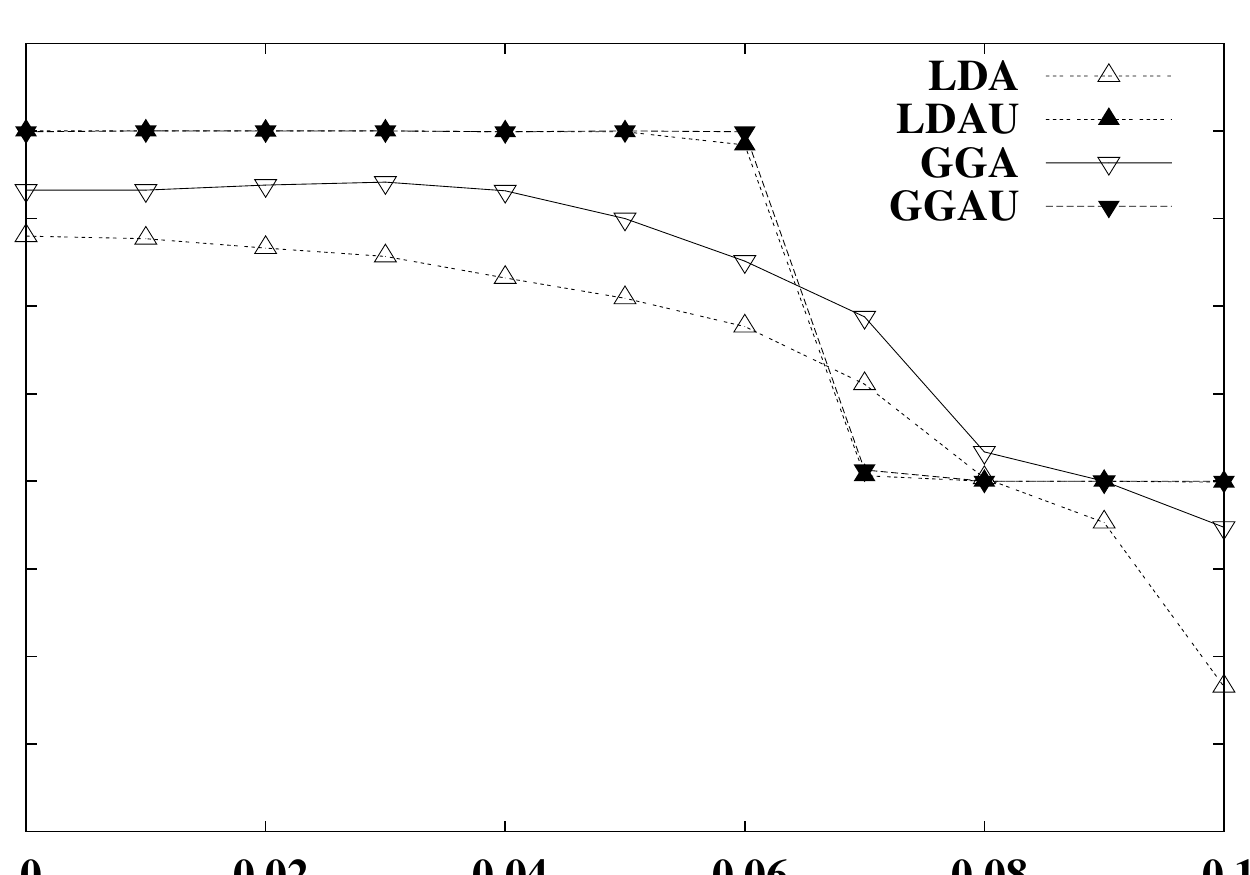}
\caption{(Color online) The calculated total magnetic moments (MM) per cell (in $\mu_{\rm{B}}$ for the JT1 system are shown in the left column. The upper panel (a) shows the MM in the AFM state whereas the lower panel (b) shows the MM in the FM state. The calculated total MM  of JT2 in the FM state are shown in the right column. Up (down) triangles represent LDA (GGA) results. Empty (filled) triangles represent results without (with) $U$.}
\label{JT1-MM}
\end{figure}

The calculated total magnetic moments per unit cell of BFO in the FM and AFM states of JT1 are shown in Fig.~\ref{JT1-MM}. The magnetic moments saturate at small displacements and are rather robust against small strains. In the small strain region the total magnetic moment per cell in the LDA$+U$/GGA$+U$ is $\sim 8.0$ $\mu_{B}$, i.e., $\sim 4.0$ $\mu_{B}$ per Fe atom, suggesting that the nominal oxidation state of Fe in this region is Fe$^{4+}$, which is consistent with the charge neutrality of BFO and agrees with experimental results.~\cite{Naoaki} However, there is a drastic change in the magnetic moments around 0.04 eV/\AA$^3$, since they tend to sharply decrease beyond this value. This is the region where the magnetic phase transition is also observed. Beyond 0.08 eV/\AA$^3$, the total magnetic moments in the FM state saturates at $\sim 4.0$ $\mu_{B}$ per cell.
In this region, LDA/GGA and LDA$+U$/GGA$+U$ almost give the same magnetic moments per unit cell.
To understand this drastic (about 50 $\%$) decrease in the value of the magnetic moment, we analyzed the Mulliken charges of the Fe atoms. The main drop in the magnetic moment comes from Fe$_{\rm I}$. This drop is due to the O$_{\rm c}$ atoms, which are displaced towards Fe$_{\rm I}$. Such displacement decreases the bond length and angle between O$_{\rm c}$ and Fe$_{\rm I}$ and strongly increases the overlap, which reduces the magnetic moment of Fe$_{\rm I}$.

The total magnetic moments per unit cell of JT1 in the AFM state (shown in Fig.~\ref{JT1-MM}) are also analyzed. In the unstrained case, BFO has zero magnetic moment which is consistent with a true AFM structure. However, when the structure is distorted, i.e., when the octahedral symmetry is lowered, BFO develops a non zero magnetic moment. Therefore, changing the atomic positions of the O atoms not only changes the local magnetic moments of the Fe atoms, but it also transforms BFO to an AFM structure with non zero magnetic moments, i.e., a FIM. It is interesting to note that starting from an AFM structure, the electronic self-consistent cycle converges to a FIM structure whenever some strain is imposed in the JT1 case. More importantly, it is also possible to see a more severe change in the total magnetic moments than that found in the FM calculations, i.e., beyond 0.07 eV/\AA$^3$, the total magnetic moment converges to $\sim -4.0$ $\mu_{B}$. Interestingly, in both FM and AFM calcul
 ations, the magnetic moments of the
unit cell converge to $\sim \left|4.0\right|$ $\mu_{B}$, which suggests that FM and AFM states are isomagnetic in this high-strain region. Note that similar considerations were also derived from the total energy calculations (See Fig.~\ref{JT1-TE})

The magnetic moments of BFO in the FM and AFM states were also examined in the JT2 case. The calculated magnetic moments in the FM states are shown in the right panel of Fig.~\ref{JT1-MM}. LDA$+U$/GGA$+U$ gives as before a magnetic moment per unit cell $\sim 8.0$ $\mu_{B}$ in the low strained region. Upon imposing strain, a drastic change in the magnetic moments is observed around $0.06$ eV/\AA$^3$. Beyond this strain the total magnetic moments converge to $\sim 4.0$ $\mu_{B}$ in the FM state, similar to the JT1 case. This drastic change in the magnetic moment is due to Fe-O$_{\rm c}$ strong bonding, as discussed in the JT1 case. In the JT2 case, both Fe atoms are moved towards the O$_{\rm c}$ atoms, which are fixed. At some strain both types of atoms bond strongly, which decreases the magnetic moments of Fe$_{\rm I}$ and Fe$_{\rm II}$. In the strained AFM case, which is more stable than the FM case, Fe$_{\rm I}$ and Fe$_{\rm II}$ have $-1.82$ and 1.82 $\mu_{B}$ magnetic moments, respectively, for $sys0.09$, where
the Fe$_{\rm I}$-O$_{\rm c}$ (Fe$_{\rm II}$-O$_{\rm c}$) bond length is $\sim 1.24$ \AA/ (2.94\,\AA). The net magnetic moment per unit cell in this case is zero, i.e., a true AFM structure. This implies that moving the Fe atoms transforms BFO from FM to AFM at large strains. This magnetic phase transition is different from that of the JT1 case, which transformed FM to FIM at large strains.

\begin{figure}[!h]
\includegraphics[width=0.4\textwidth]{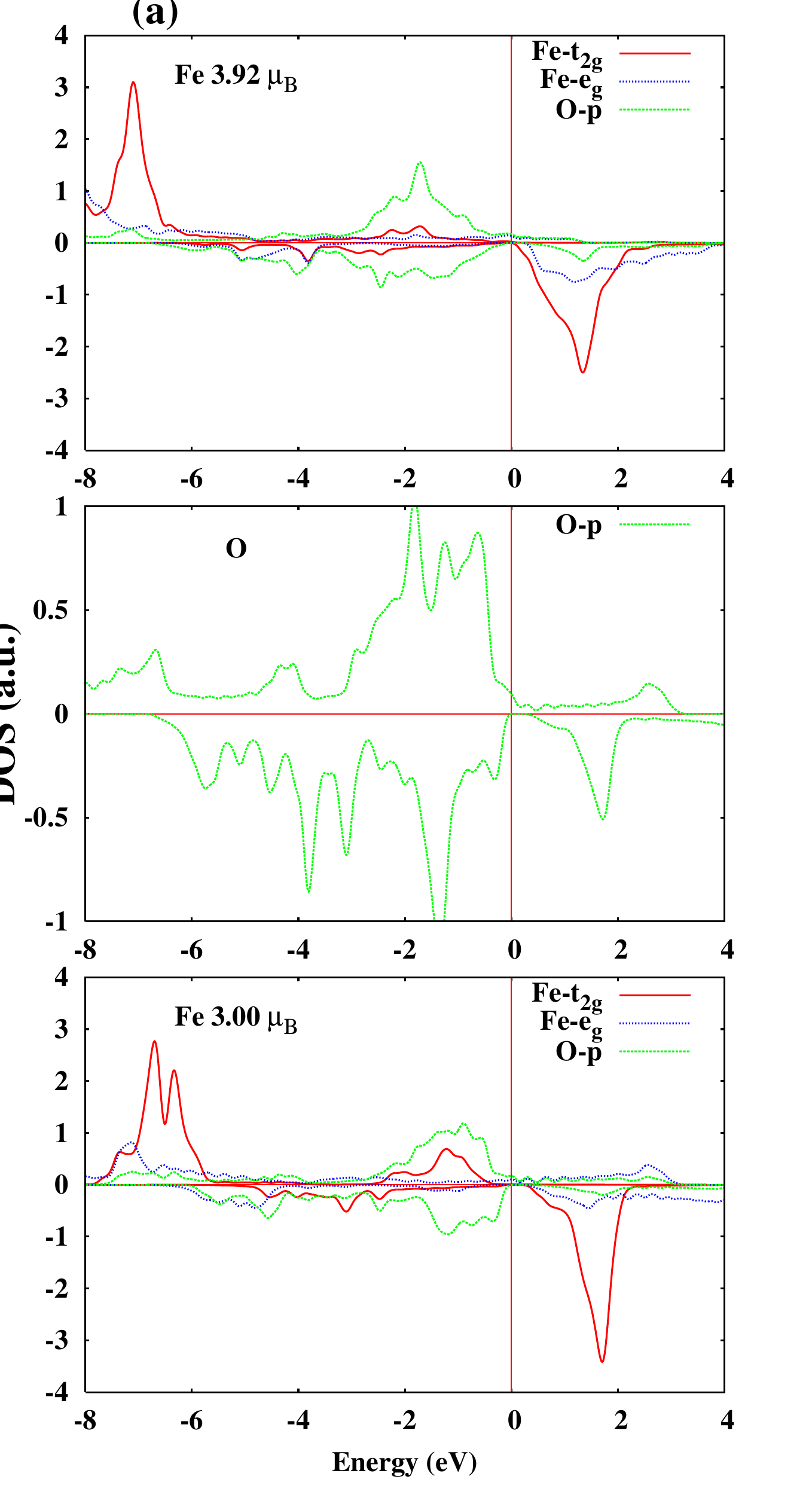}
\includegraphics[width=0.4\textwidth]{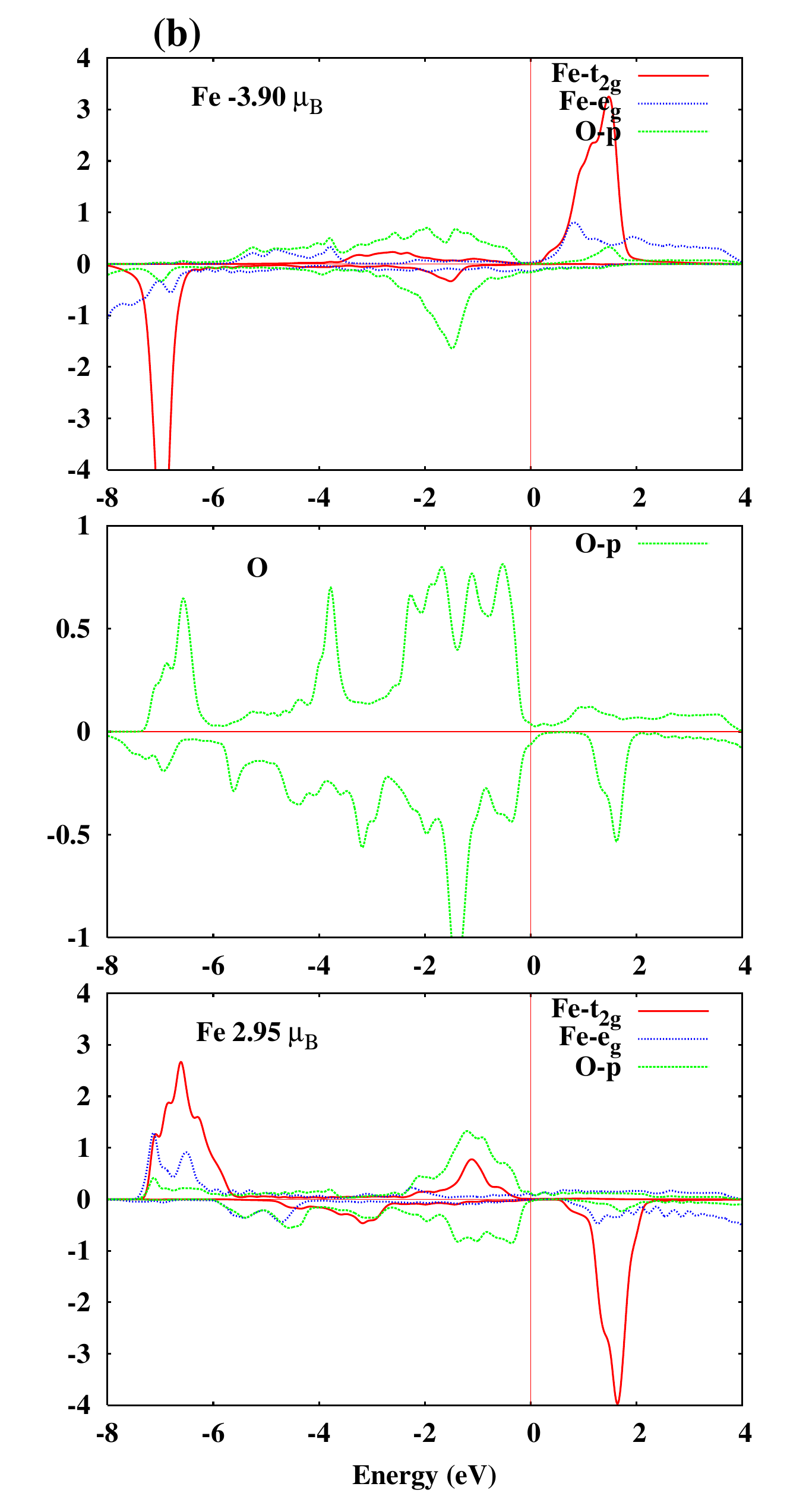}
\caption{(Color online)
GGA$+U$ calculated PDOS (in arbitrary units) in the FM (left panels, labeled as (a)) and AFM (right panels, labeled as (b)) states for JT1, $sys03$. Solid (red), dotted (blue), and dashed (green) lines represent Fe-$t_{2g}$, Fe-$t_{eg}$, and O-$p$ orbitals, respectively. The local magnetic moments of the Fe atoms are also shown in each panel. The Fermi energy is set to zero.}
\label{DOS-Sys3}
\end{figure}

To elucidate the origin of these magnetic phase transitions and the reduction of the magnetic moments beyond some critical strain, we focus on the effect of distortions on the electronic structure of BFO. As we previously showed, pristine BFO has a half metallic electronic structure where $t_{2g}$ and $e_{g}$ states are separated from each other due to a crystal field splitting (see Fig.\ref{DOSFM}). The calculated DOS for the JT1 case in the FM and AFM states for a strain of 0.03 eV/\AA$^3$, which is very close to the FM-AFM/FIM transition, are given in Fig.~\ref{DOS-Sys3}. As can be seen, the electronic structure remains half metallic but now the low lying $t_{2g}$ band is splitted into two energy states, particularly in the Fe$_{\rm I}$ $d$ band. A broad majority $t_{2g}$ band is also formed just below the Fermi energy, which shows a strong bonding with the O$_{\rm c}$ atom connecting Fe$_{\rm I}$ and Fe$_{\rm II}$. Both Fe atoms have different local magnetic moments
and exchange splittings, which implies that strain changes the local symmetry and distorts differently the electronic structure of each atom. Similar behavior can be seen in the AFM DOS as well.

\begin{figure}[!h]
\includegraphics[width=0.4\textwidth]{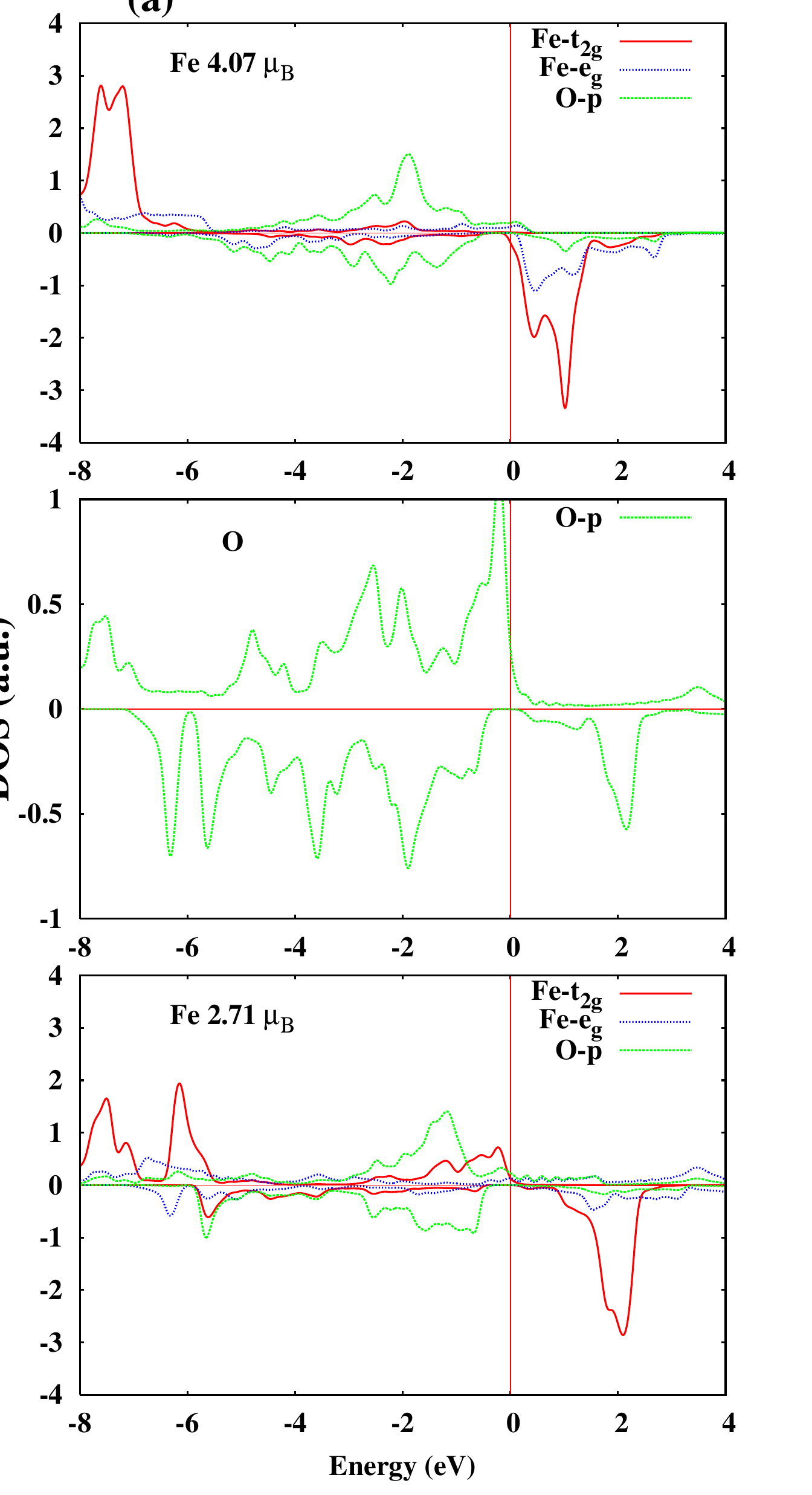}
\includegraphics[width=0.4\textwidth]{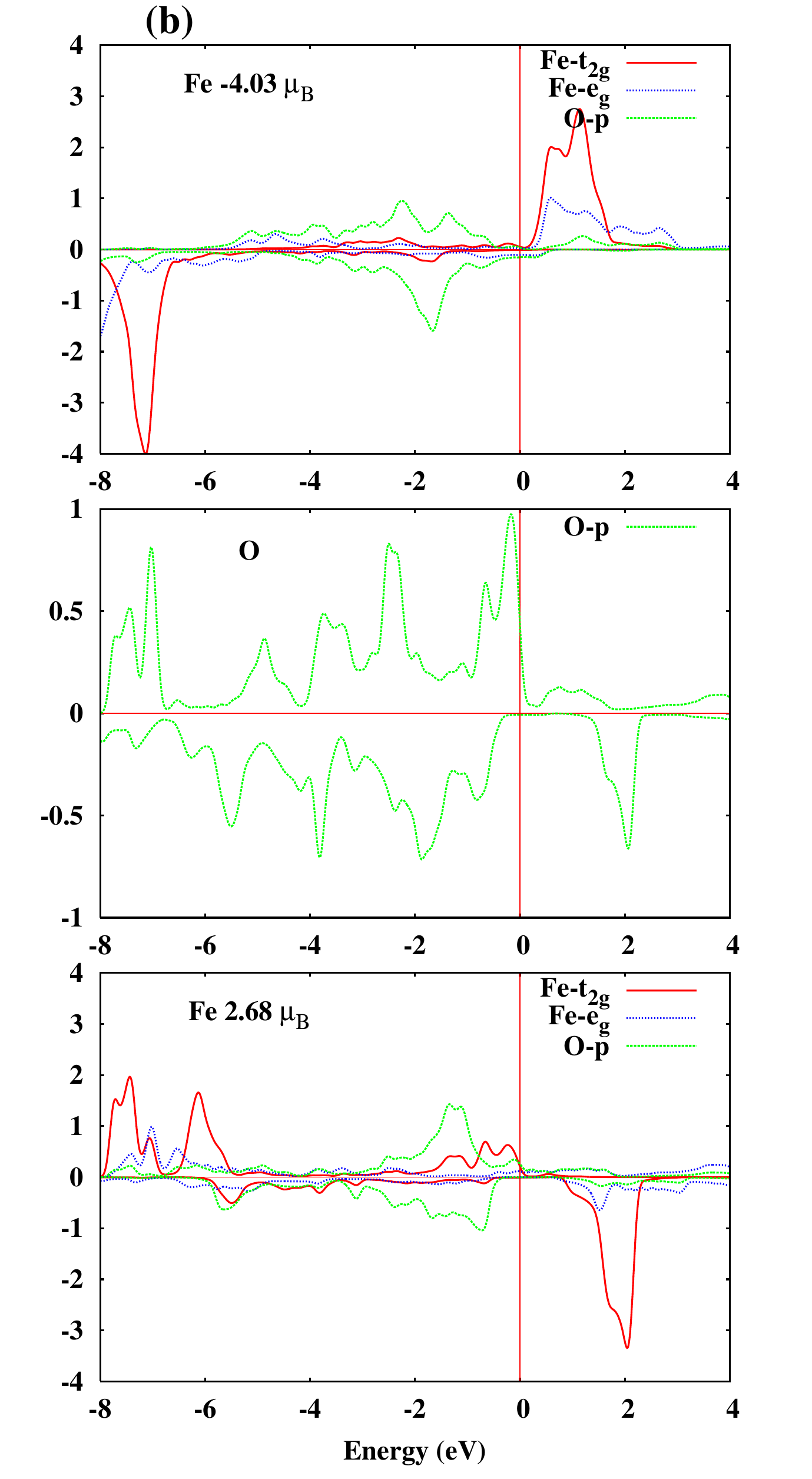}
\caption{(Color online)
GGA$+U$ calculated PDOS (in arbitrary units) in the FM (left panels,labeled as (a)) and AFM (right panels, labeled as (b)) states for JT1, $sys05$. Solid (red), dotted (blue), and dashed (green) lines represent Fe-$t_{2g}$, Fe-$t_{eg}$, and O-$p$ orbitals, respectively. The local magnetic moments of the Fe atoms are also shown in each panel. The Fermi energy is set to zero.}
\label{DOS-Sys5}
\end{figure}

\begin{figure}[!h]
\includegraphics[width=0.4\textwidth]{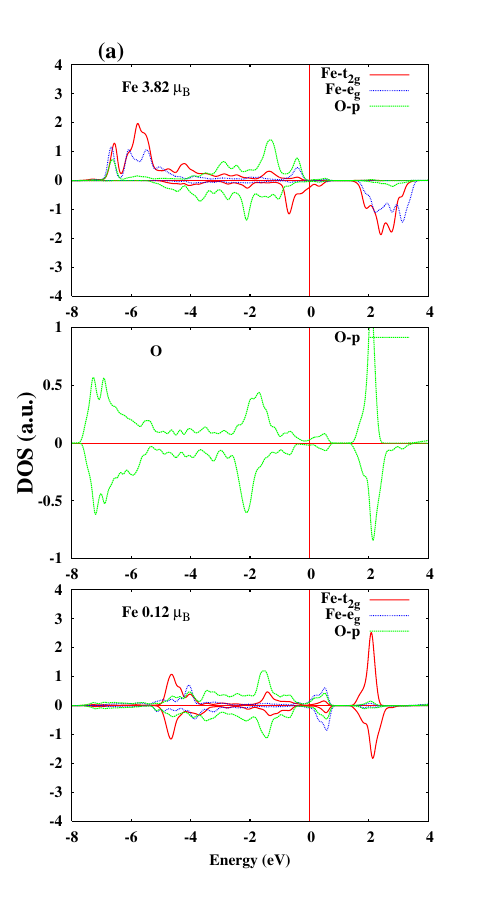}
\includegraphics[width=0.4\textwidth]{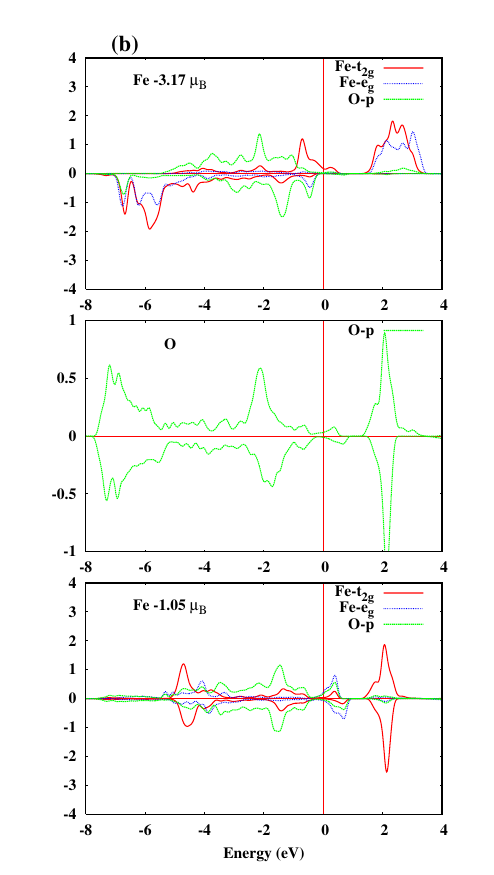}
\caption{(Color online)
GGA$+U$ calculated PDOS (in arbitrary units) in the FM (left panels,labeled as (a)) and AFM (right panels, labeled as (b)) states for JT1, $sys09$. Solid (red), dotted (blue), and dashed (green) lines represent Fe-$t_{2g}$, Fe-$t_{eg}$, and O-$p$ orbitals, respectively. The local magnetic moments of the Fe atoms are also shown in each panel. The Fermi energy is set to zero.}
\label{DOS_Sys9}
\end{figure}

We consider now the electronic structure in the strained region, where AFM/FIM is the ground state structure, to have a clear understanding of such magnetic phase transition. Figure \ref{DOS-Sys5} shows the electronic structure of JT1 BFO under 0.05 eV/\AA$^3$ ($sys0.05$) in the FM and AFM states. The most noticeable feature of this electronic structure is a clear splitting of the low lying Fe $t_{2g}$ band, in both the FM and AFM states. At this strain, the JT distortion is almost completed at the Fe$_{\rm I}$ site, while Fe$_{\rm II}$ will show a similar JT distortion at higher strains. The FM-AFM/FIM magnetic phase transition is mainly caused by such JT distortion at the Fe$_{\rm I}$ site, which destabilizes the FM ground state and reduces the local magnetic moment, as compared with the pristine and low strained systems. The oxygen PDOS associated to O$_{\rm c}$ is also significantly changed, and becomes partially occupied.
Therefore, the FM state becomes unstable when the PDOS of O$_{\rm c}$ in the spin-up state touches the Fermi energy. At larger strains the O$_{\rm c}$ PDOS further crosses the Fermi energy, which stabilizes the AFM/FIM state (see also Fig.~\ref{DOS_Sys9}). 

Figure~\ref{DOS_Sys9} shows the PDOS in the FM and AFM states of JT1 $sys0.09$. Compared with the unstrained BFO in the FM state, the $t_{2g}$ states of Fe are shifted now towards the Fermi energy. There are also unoccupied $d$ states which were occupied in the unstrained case. The exchange splitting of the Fe$_{\rm I}$ $d$ band significantly decreases, which reduces its magnetic moment to 0.12 $\mu_{B}$ in the FM state. However, Fe$_{\rm II}$ still carries a large magnetic moment. The local moments of both Fe atoms are therefore different (see also their PDOS), which implies the presence of different crystallographic environments.~\cite{GR-MnAs2} In such different local environments the Fe atoms generate a FIM-type behavior, similar to FIM Fe$_{2}$O$_{3}$. The band structure in the AFM case is similar and shows that Fe$_{\rm I}$ and Fe$_{\rm II}$ are not coupling antiferromagnetically but ferrimagnetically, which is consistent with the local magnetic moments of Fe$_{\rm I}$ and Fe$_
 {\rm II}$. The DOS also demonstrates the existence of an electronic phase transition (half metal to metallic) at large strains. The magnetic phase transition is therefore accompanied by an electronic phase transition at large strains, mainly contributed by $t_{2g}$ and O $p$ orbitals.

\begin{figure}[htb]
\centering
\includegraphics[width=0.2\textwidth,width=0.2\textwidth ]{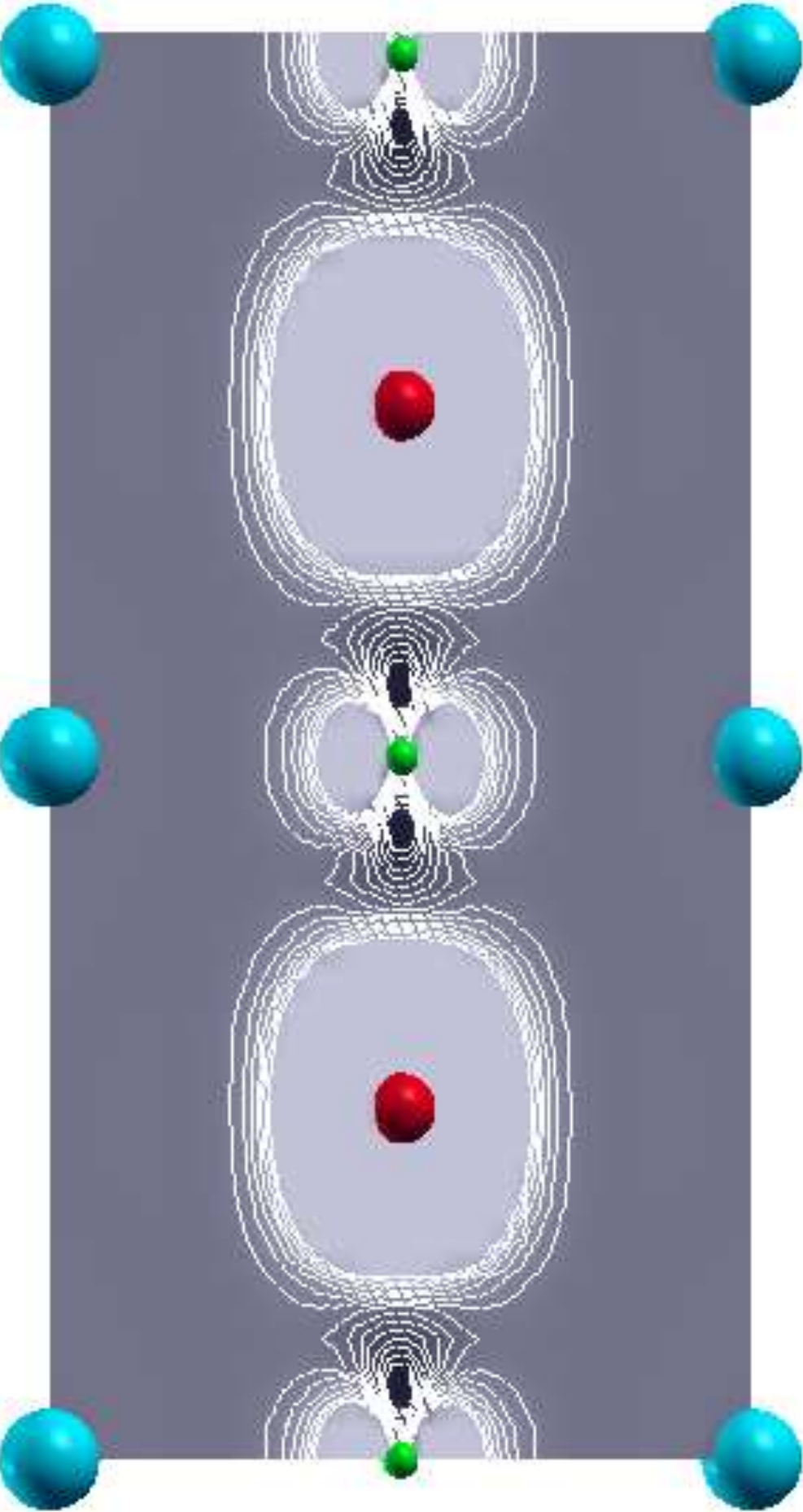}\hspace{2cm}
\includegraphics[width=0.2\textwidth,width=0.2\textwidth ]{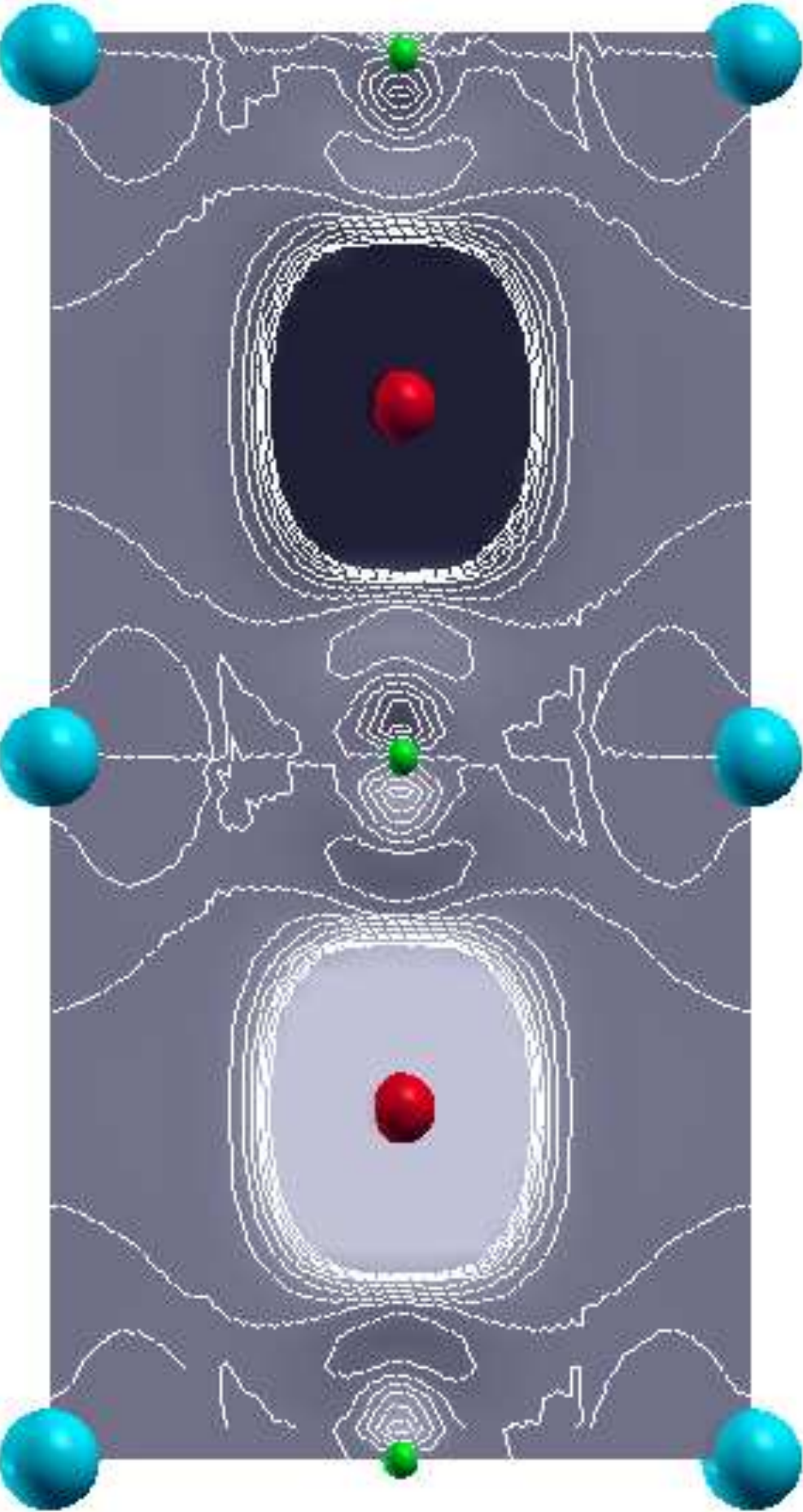}\\
\includegraphics[width=0.2\textwidth,width=0.2\textwidth ]{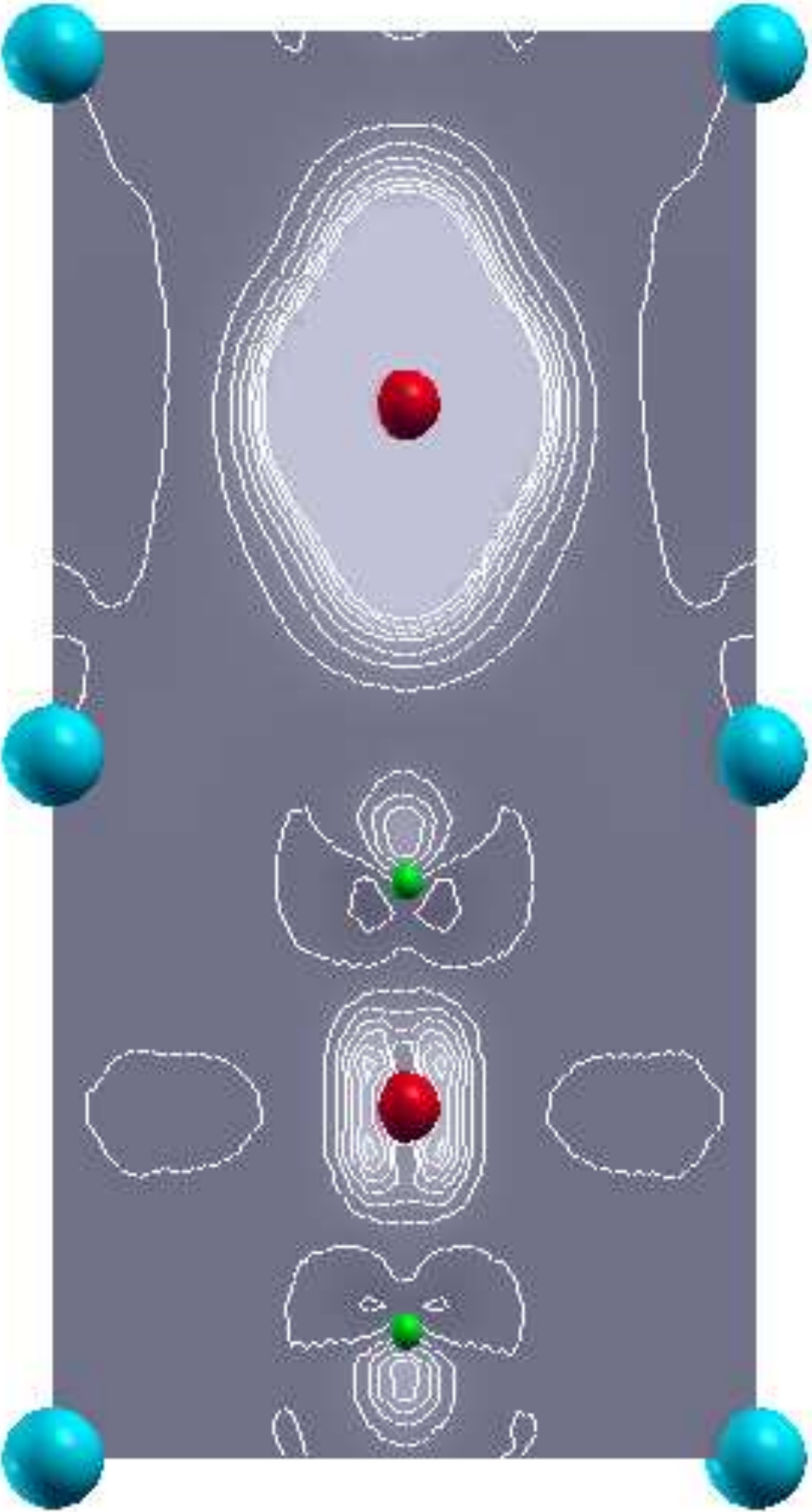}\hspace{2cm}
\includegraphics[width=0.2\textwidth,width=0.2\textwidth ]{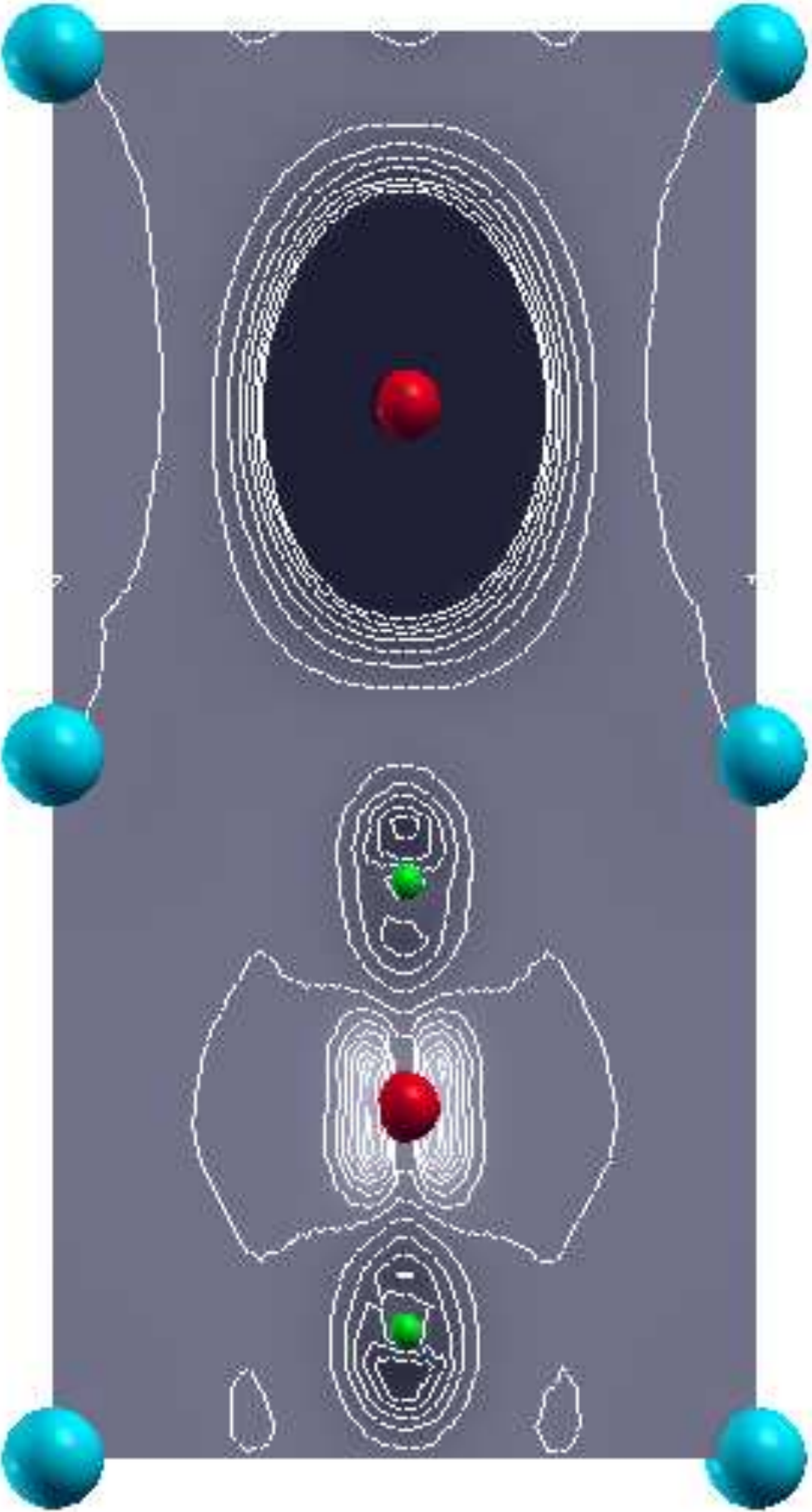}\\
\includegraphics[width=0.2\textwidth,width=0.2\textwidth ]{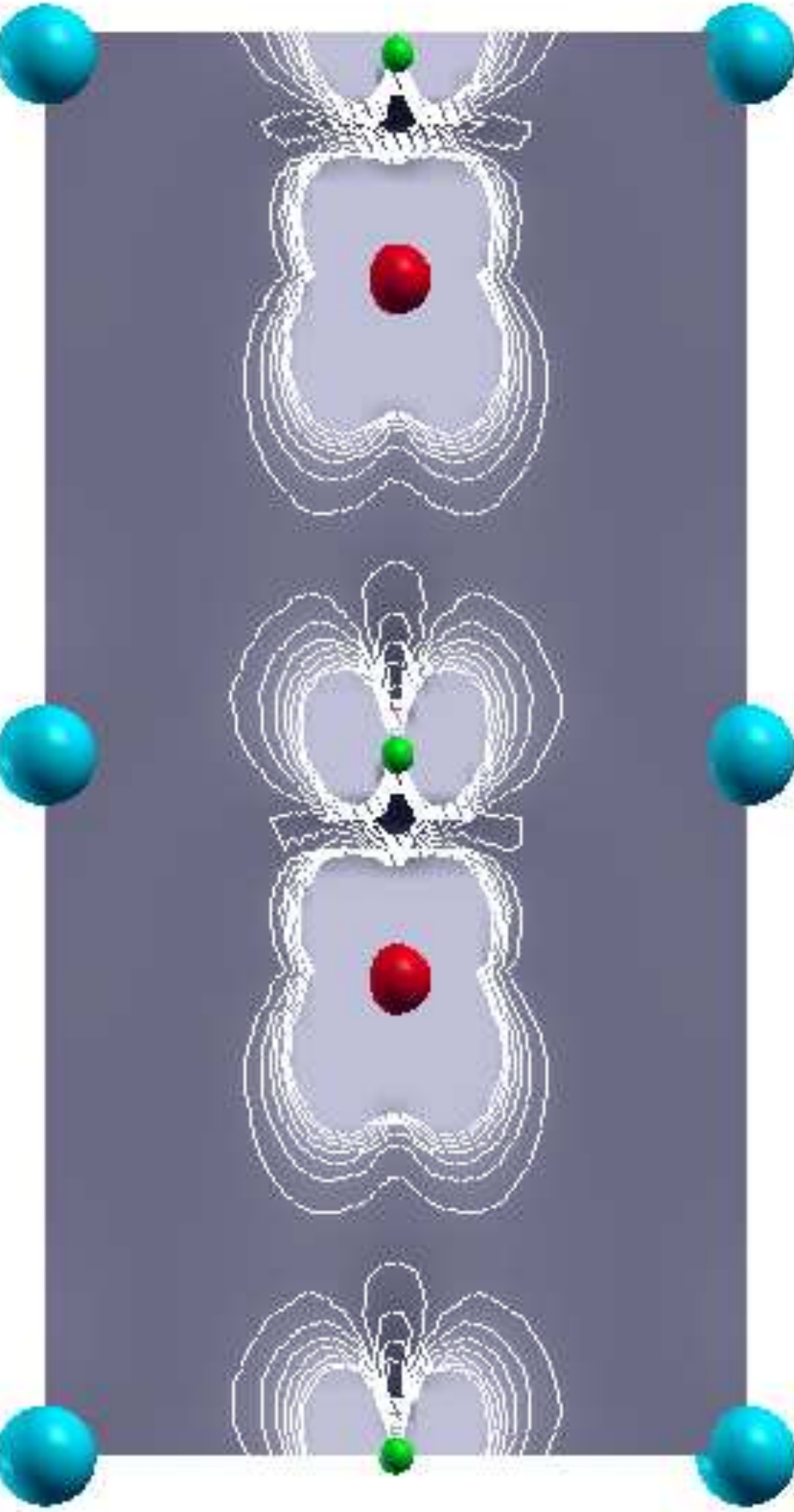}\hspace{2cm}
\includegraphics[width=0.2\textwidth,width=0.2\textwidth ]{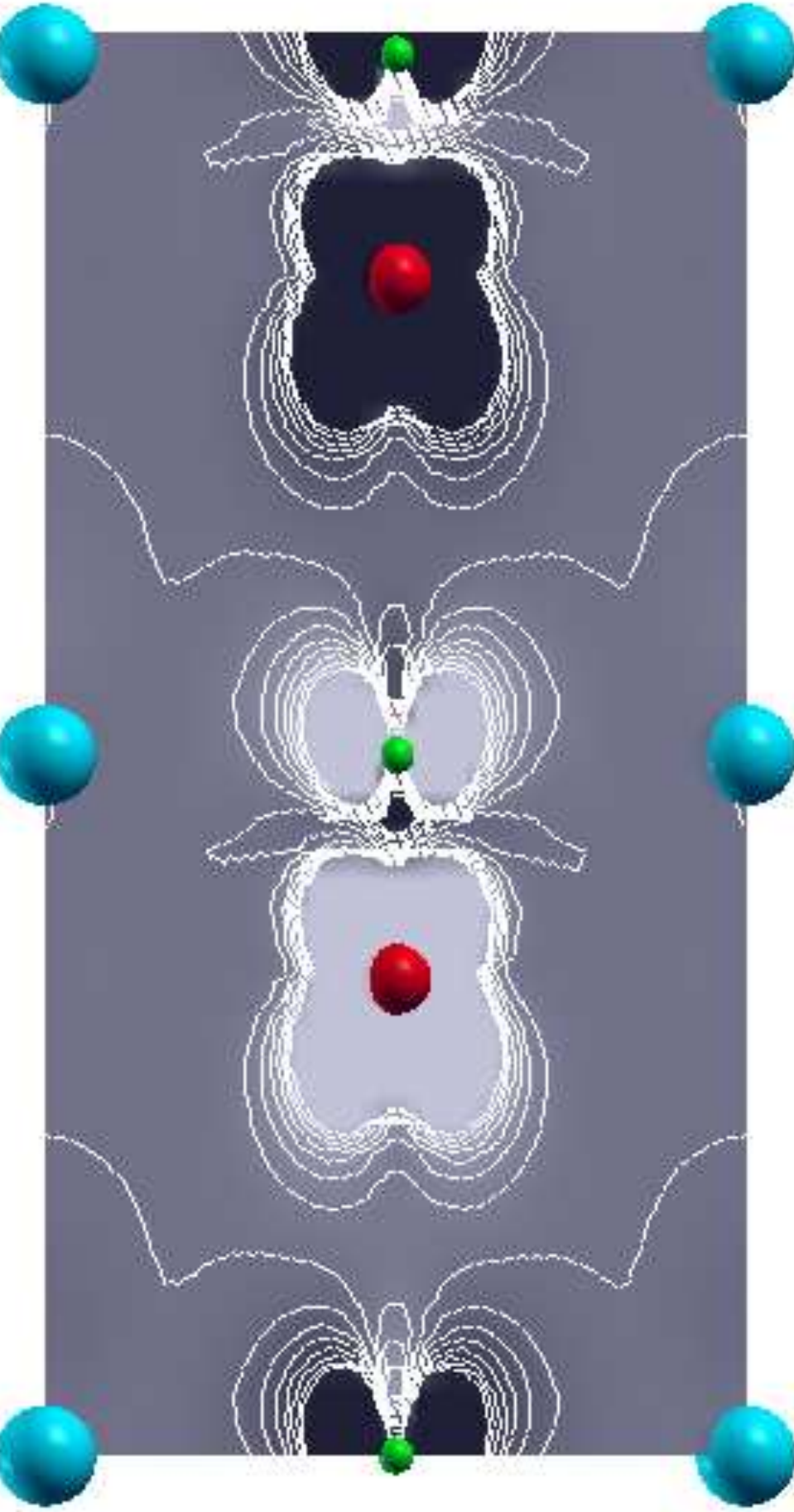}\\
\includegraphics[width=0.2\textwidth,width=0.2\textwidth ]{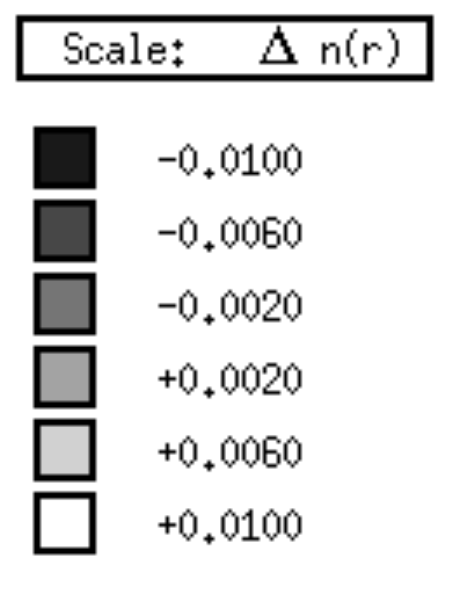}
\caption{(Color online) Spin densities in the FM (left) and AFM (right) states for $sys0$ (top)and $sys09$ (middle)of JT1 and ($sys09$) of JT2. Cyan, red, and green balls represent Ba, Fe and O atoms, respectively. The scale is also shown at the bottom of the figure.}
\label{spin-dos}
\end{figure}

To shed more light on the mechanism of the phase transition we look now into the spin densities of different systems in the FM and AFM states in the (110) plane. This analysis will further help us to understand the origin of magnetic phase transitions and the reduction of magnetic moments at large strains. Figure~\ref{spin-dos} shows the spin densities of JT1 $sys00$, $sys009$ and JT2 $sys009$. The spin denisty of $sys00$ exhibits a large spin polarization at O$_{\rm c}$, both in the FM and AFM states and a large spin density at the Fe sites. Also, a clear anitiferromagnetic coupling between Fe$_{\rm I}$ and Fe$_{\rm II}$ can be seen in the AFM state. Notice that as we increase the strain the magnetic moment at the Fe$_{\rm I}$ site starts decreasing because the O$_{\rm c}$ atom connecting Fe$_{\rm I}$ and Fe$_{\rm II}$ moves towards Fe$_{\rm I}$, which decreases its magnetic moment. This behavior can be seen in Figure~\ref{spin-dos}, where Fe
$_{\rm II}$ has a large spin polarization ,whereas Fe$_{\rm I}$ has a smaller magnetic moment of $\sim 0.12$ $\mu_{B}$ ($\sim -1.05$ $\mu_{B}$) in the FM (AFM) state. Notice also the spin polarization at the O$_{\rm c}$ site is reduced. This behavior shows again that the FM-AFM/FIM transition is mainly governed by strong Fe$_{\rm I}$-O$_{\rm c}$ bondings. In the JT2 $sys009$ the spin densities of both Fe atoms decrease again due to the strong overlap with the O$_{\rm c}$ atom but their shape is indistinguishable, which is expected since the local environments are the same.

The origin of the FM-AFM/FIM phase transition can be discussed in terms of SE and DE models. Generally, compressing a lattice parameter $a$ leads to the increase of the hopping integral $t$, represented by $pd \sigma$. In this particular case $t$ personifies the orbital hybridization between O ${p}$ and Fe$\,3{d}$. Since DE and SE energies are proportional to $t$ and $t^{4}$, respectively, increasing $t$ enhances SE as compared with DE.\cite{{Most}} In our case the lattice constant $a$ of BFO  is fixed at its optimized value, and we only changed the bond angle/length between the atoms in such a way that it strongly modifies the electronic structure. As we have shown (see all DOS figures) when the strain increases, the hybridization between the Fe and O orbitals also increases. At a large strains $t$ becomes very strong and BFO transitions to AFM/FIM. This shows that SE is enhanced by strain, since the oxygen band width increases and favors AFM/FIM. On the other hand, at low strain
the overlap  between Fe $3d$ and O $2p$ is relatively weak and DE wins, which compels BFO to be in the FM state. This competition between DE and SE is consistent with our total energy calculations, i.e., FM vs. AFM states. At large strain the energy difference $J$ increases, suggesting that $t$ also increases, which agrees with the DOS figures. In such competition, the magnetic moments decrease due to the strong $t$.
Changing $a$ can also change the magnetic structure of BFO, i.e., A-type or G-type helical spin structures. In particular, decreasing $a$ produces a G-type structure.~\cite{LiPRB2012} Such magnetic structural changes are again mainly caused by $t$. In the light of the phase diagrams  presented in Ref.\onlinecite{Most}(Fig.3), where it is shown that the transition from FM to a helical magnetic (HM) phase transition depends on $pd\sigma$ and oxygen-oxygen hoping amplitudes $t_{pp}$, it is possible to infer that large $t_{pp}$ is also important for the stabilization of the HM state. In our case, straining BFO produces AFM or FIM magnetic states, depending on the nature of the strain. Therefore, we believe that the experimentally observed magnetic phase transition at high pressure or external magnetic fields\cite{Naoaki} may distort the local environment of the Fe and O atoms in BFO and such distortion can favor FM-FIM/AFM magnetic phase transitions.

\section{Summary}
Cubic BaFeO$_{3}$ (BFO) was investigated using density functional theory (DFT) within the generalized gradient approximation (GGA) and the local density approximation (LDA). The effect of on site Coulomb interaction $U$ was also considered. We showed that BFO has a ferromagnetic (FM) ground state in all cases. This FM ground state is in agreement with recent experimental work. The electronic and magnetic structure of BFO was modified by two different types of Jahn-Teller (JT) distortions, denoted as JT1 and JT2. We showed that JT frustrates the magnetic structure of BFO and induces FM to antiferromagnetic (AFM) and FM to ferrimagnetic (FIM) transitions. Such magnetic phase transitions are governed by the displacements of the O and Fe atoms from their ideal positions. The DFT and DFT$+U$ calculations reveled that a larger JT distortion is needed for the FM to AFM transition as compared to the FM to FIM transition. The Coulomb interaction mainly shifted the magnetic phase transition to
  higher strains. The JT distortions also reduced the magnetic moments of the Fe atoms. The mechanisms of the transitions were elucidated using partial densities of states and spin density contours. These results suggest that different magnetic phase transitions can be induced in BFO by different types of strain.

\section*{ACKNOWLEDGMENTS}
We are greatful to Erjun Kan for fruitful discussions. G. R. acknowledges the cluster facilities of NCP, Pakistan.
J.M.M. acknowledges computational resources provided by CESUP-UFRGS, Brazil.
R.F. acknowledges financial support through a Severo-Ochoa grant (Consejer\'{\i}a
de Educaci\'on, Principado de Asturias). V.M.G.S. thanks the Spanish Ministerio
de Econom\'{\i}a y Competitividad for a Ram\'on y Cajal fellowship (RYC-2010-06053).


\newpage

\newpage

\newpage

\newpage

\newpage

\newpage

\newpage

\newpage

\end{document}